\documentclass[software,article,accept,pdftex,oneauthor]{Definitions/mdpi}

\providecommand{\hreflink}[1]{}
\providecommand{\TitleCitation}[1]{}
\providecommand{\AuthorCitation}[1]{}

\renewcommand{\highlighting}[1]{#1}
\def\hl#1{#1}

\usepackage[utf8]{inputenc}
\usepackage{lipsum}
\usepackage{xcolor}
\usepackage{tabularx}
\usepackage{booktabs}
\usepackage{multirow}
\usepackage{graphicx}
\usepackage{amsmath}
\usepackage{listings}
\usepackage{makecell} 

\firstpage{1}
\pubvolume{5}
\issuenum{2}
\articlenumber{26}
\pubyear{2026}
\copyrightyear{2026}
\externaleditor{Tadashi Dohi, Junjun Zheng and Xiao-Yi Zhang}
\datereceived{13 May 2026}
\daterevised{9 June 2026}
\dateaccepted{16 June 2026}
\datepublished{18 June 2026}

\Title{\textls[-15]{Iterative Audit Convergence in LLM-Managed Multi-Agent Systems: A Case Study in Prompt-Engineering Quality Assurance}}

\Author{Elias Calboreanu \orcidA}

\AuthorNames{Elias Calboreanu}

\address[1]{Swift AI Lab, The Swift Group, LLC, Reston, VA 20190, USA; ecalboreanu@theswiftgroup.com}

\abstract{\highlighting{Prompt} %
 specifications for multi-agent large language model (LLM) systems carry data contracts and integration logic across interdependent files but are rarely subjected to structured-inspection rigor. We report a single-system case study of iterative, agent-driven auditing applied to AEGIS (Autonomous Engineering Governance and Intelligence System), a seven-lane production pipeline whose 7152-line specification surface was audited across nine rounds, surfacing 51 consistency defects (per-round counts of 15, 8, 12, 2, 8, 1, 4, 1, 0). We present a seven-category post hoc taxonomy with explicit coding rules, non-monotonic convergence consistent with cascading edits and audit-scope expansion, and a locked audit protocol. We further report two partial replications on a public synthetic mini-specification: a cross-LLM panel of four frontier vendors (OpenAI, Anthropic, Google, xAI; 12 traces; multi-vendor union detects all five seeded defects) and an inter-rater reliability check on a stratified subsample (Cohen's $\kappa$ = 0.80 on category, 0.46 on severity). The full reproducibility bundle accompanies the submission.}


\begin{document}

\section{Introduction}
\label{sec:introduction}
\unskip

\subsection{Motivation}

Prompt engineering for multi-agent systems presents a quality assurance challenge distinct from traditional software testing.~When an LLM-orchestrated pipeline spans multiple autonomous lanes, each with its own behavioral specification, data contracts, and~integration points, the~prompt layer becomes the system's source of truth. Errors in prompt specifications can propagate as runtime failures: wrong field names may break data handoffs, stale references can trigger invalid API calls, and~missing labels can corrupt downstream classification. The~empirical evidence presented in this paper is consistent with such failure modes but does not establish their relative~frequency.

However, prompt specifications are rarely subjected to the rigor applied to source code. They are authored in natural language, reviewed informally, and~updated incrementally. When a design change touches one document, ripple effects across other documents are easy to miss.~Software inspection research has long established that structured review catches defects that testing alone may miss~\cite{fagan1976design,boehm2001defect,weinberg1984reviews}.~The~challenge specific to multi-agent LLM systems is that the inspector is most naturally an LLM as well: human reviewers do not scale to seven interlinked lane specifications updated weekly, while LLM-driven walkthroughs can. That choice introduces a same-LLM-family concern---an inspector from the same model family as the author can share blind spots---which structures the design of this study and is taken up directly in the cross-LLM partial replication of Section~\ref{subsec:cross-llm}.

\textbf{\hl{Objective:}%
} The single main objective of this paper is to characterize, on~one production system, what iterative LLM-driven auditing of a prompt-specification surface looks like in practice: which defect classes surface only across iterations, how convergence behaves, and~what protocol parameters matter for replication.~We pair the case study with two partial replications on a public synthetic mini-specification---a cross-LLM panel of four frontier vendors and an inter-rater reliability check---to bound the most-flagged threats to validity (a single LLM family auditing the system and single-coder taxonomy) before any generalization is~attempted.

\subsection{Research~Questions}

The research questions are stated descriptively rather than causally:

\begin{itemize}
\item \textbf{\hl{RQ1.}} What defect classes remained after the initial audit pass and surfaced only through later iterative expanded-scope rounds in this case study?
\item \textbf{\hl{RQ2.}} What convergence behavior was observed, and~how many rounds were required for this 7152-line, eight-document specification surface?
\end{itemize}

\subsection{Contributions}

This paper offers four contributions, each scoped to the case study and its \linebreak  partial replications:

\begin{itemize}
\item \textls[-15]{\textbf{\hl{C1 (taxonomy).}} A seven-category post hoc taxonomy of the 51 surfaced defects, with~category definitions, coding rules, and~one representative example per category (Section~\ref{sec:results} and \highlighting{Appendix~\ref{appendix:b}}%
). Percentages are descriptive; no inferential claims are made.}
\item \textbf{\hl{C2 (observed convergence).}} Per-round defect counts with Wilson 95\% confidence intervals, a~Mann--Kendall trend statistic on the convergence series, and~a category-by-round cross-tabulation showing observed non-monotonic intermediate behavior consistent with cascading edits and audit-protocol maturation.~We do not claim that the curve is predictable in advance; we report what was observed under explicitly \linebreak  stated confounds.
\item \textbf{\hl{C3 (audit protocol distilled from the study).}} An audit protocol distilled from the case study, with~the final locked checklist, prompt template, finding schema, convergence criterion, and~context-loading chronology released as a reproducibility appendix (Appendices~\ref{appendix:a} and~\ref{appendix:c}). We distinguish \emph{\hl{protocol used during the study}%
}, which co-evolved with discovery, from~\emph{\hl{protocol recommended after the study}}.
\item \textbf{\hl{C4 (partial replications bounding the central threats).}} On a released public synthetic mini-specification, a~four-vendor cross-LLM audit panel (12 traces from OpenAI, Anthropic, Google, and~xAI; Section~\ref{subsec:cross-llm}) and an inter-rater reliability check on a stratified 18-defect subsample (Cohen's $\kappa = 0.80$ on category, $0.46$ on severity; Section~\ref{subsec:irr}). Both are partial replications, not multi-system or human-IRR validations, and~are \linebreak  scoped accordingly.
\end{itemize}

We do not claim that single-pass full-scope review is universally insufficient---only that single-file review cannot, by~construction, detect cross-file defects, and~that one-pass termination would have left later-discovered defects unresolved under this study's protocol. A~single-pass full-scope control was not~run.

\section{Background and Related~Work}
\label{sec:background}
\unskip

\subsection{Software Inspection and Structured~Review}

Fagan-style code inspection~\cite{fagan1976design} established that structured review detects defects that testing alone may miss.~Boehm and Basili's ``Software Defect Reduction Top 10 List''~\cite{boehm2001defect} reaffirmed this in 2001 with broader empirical evidence, identifying peer reviews and inspections as cost-effective defect-reduction mechanisms across multiple project contexts. Weinberg and Freedman~\cite{weinberg1984reviews} articulated lighter-weight walkthroughs and reviews, observing that effectiveness depends on checklist quality and reviewer expertise without requiring the full Fagan moderator/reader/recorder/inspector role~separation.

Our audit protocol adapts the \emph{\hl{walkthrough}} tradition rather than full Fagan inspection. We did not have multiple human inspectors with separate role assignments; we had one LLM family acting as the inspector against a checklist authored by the same family. The~locked checklist in Appendix~\ref{appendix:a} serves as the inspection guide, the~sub-agent serves as the inspector, and~the convergence criterion replaces the single-pass exit gate. The~key limitation, that the inspector belongs to the same LLM family that authored the specifications, is documented in Section~\ref{subsec:threats}.

\subsection{Specification Drift, Aging, and~Traceability}

Parnas~\cite{parnas1994software} identified software aging as the phenomenon by which documentation accuracy degrades as the system that it describes evolves. Gotel and Finkelstein~\cite{gotel1994analysis} analyzed the requirements traceability problem, finding that cross-artifact consistency is the most labor-intensive and error-prone aspect of requirements management.~Our findings are consistent with both observations in the prompt-specification domain: 12 of the 51 defects (23.5\%) were stale Jira references and 5 (9.8\%) were cross-lane schema~mismatches.

\subsection{LLM Self-Evaluation and~Self-Refinement}

Three nearby lines of work all operate \emph{\hl{within}} an LLM's context window; our case study operates \emph{\hl{across}} sessions on persistent file-based artifacts. Self-consistency~\cite{wang2023self} samples multiple reasoning paths and selects the most consistent answer to a single prompt. Self-refinement~\cite{madaan2023self} iterates a critique-and-revise loop on a single completion within one LLM. Lubos~et~al.~\cite{lubos2024llm_requirements_qa} evaluated LLMs as requirements-QA checkers against ISO/IEC/IEEE 29148~\cite{iso29148}-style quality characteristics applied to one requirement at a time. Each of these methods asks the following question: \emph{\hl{Is this completion or this requirement consistent with itself or with a rubric, right now?}} The present study asks a different question: \emph{\hl{Across many persistent files and multiple sessions, can iterative audit-and-fix drive a specification surface to quiescence at all, and~what does the convergence path look like?}} The persistence of file-based state changes the convergence dynamics: drift can accumulate across rounds, fixes can cascade into new defects, and~a single inspector's blind spots persist across iterations unless cross-vendor evidence is introduced. Patil~\cite{patil2025llm_sqa} and Akshathala~\cite{akshathala2025beyond_task_completion} made the parallel point for LLM-based software quality assurance: final-pass success on its own is~insufficient.

\subsection{Recent LLM-Based SE and Agentic Quality~Assurance}

He, Treude, and~Lo~\cite{he2024llm_multiagent} surveyed LLM-based multi-agent systems across the software development lifecycle and identified robustness and synergy gaps. Cemri~et~al.~\cite{cemri2025multiagent_fail} proposed a 14-failure-mode taxonomy for multi-agent LLM systems with human-annotated inter-rater agreement (\(\kappa = 0.88\)); their work directly motivates Appendix~\ref{appendix:b}'s coding rules. Lubos~et~al.~\cite{lubos2024llm_requirements_qa} evaluated LLMs for requirements quality assurance against \hl{ISO 29148}%
-style requirement-quality~characteristics.

Chen~et~al.~\cite{chen2025promptware} argued that prompts are software artifacts requiring lifecycle methods (the \emph{\hl{promptware}} position). Patil~\cite{patil2025llm_sqa} and the agentic-AI assessment-framework literature~\cite{akshathala2025beyond_task_completion} are consistent with the position that final-pass success on its own is an inadequate quality signal.~Repository-level agentic code reasoning benchmarking~\cite{li2026repo_reason} provides a parallel for cross-file consistency challenges.~Yang~et~al.'s SWE-agent~\cite{yang2024sweagent} establishes a leading framework for LLM-driven single-agent repository interaction; AgentBench~\cite{liu2024agentbench} provides standardized evaluation of LLMs-as-agents in synthetic environments. The~present case study complements both lines of work by reporting production-deployment evidence rather than benchmark performance. MetaGPT~\cite{hong2023metagpt} operationalizes standard operating procedures for multi-agent collaboration; AEGIS's seven-lane structure is a domain-specialized instance of that pattern. Naqvi, Baqar, and~Mohammad~\cite{naqvi2026agentic_testing} contrasted iterative closed-loop multi-agent testing against static single-shot test~generation.

\subsection{Companion~Work}

The companion preprint~\cite{calboreanu2026aegis_companion} addresses the runtime behavior of the same AEGIS pipeline, including STRIDE-based adversarial testing and FMEA-based safety analysis. The~51 specification defects reported here are distinct from the 51 STRIDE-categorized adversarial code findings reported in the companion: those are P0--P3 priority code vulnerabilities (lock handling, race conditions, subprocess management) found by a two-person human review panel across three injection rounds; ours are spec-consistency defects (version drift, stale Jira refs, schema mismatches) found by Claude sub-agents across nine audit rounds. The~numerical coincidence is genuine but the populations are distinct in source, method, and~content.

This paper exercises the prompt-specification layer that configures AEGIS; it does not validate the upstream Context Engineering~\cite{calboreanu2026context}, MANDATE~\cite{calboreanu2026mandate}, LATTICE~\cite{calboreanu2026lattice}, or~TRACE~\cite{calboreanu2026trace} frameworks themselves. Those frameworks provide system context~only.

\section{\highlighting{Materials and~Methods} %
}
\label{sec:methods}
\unskip

\subsection{System Under Test: AEGIS}

AEGIS (Autonomous Engineering Governance and Intelligence System) is a seven-lane agentic automation pipeline operating against a Jira backlog that had grown to 2100+ tickets as of April 2026, up~from the companion paper's approximately 1602-row consolidated backlog snapshot reported in~\cite{calboreanu2026aegis_companion}. Each lane is a Claude agent governed by a \texttt{\hl{PROMPT.md}%
} file that defines the following:

\begin{itemize}
\item Behavioral rules and constraints;
\item Data schemas for inter-lane communication;
\item Model Context Protocol (MCP)~\cite{anthropic2024mcp} tool permissions and rate limits;
\item Jira transition contracts (status IDs, label conventions);
\item Worker pool configuration (parallelism, isolation);
\item Quality gates.
\end{itemize}

The seven lanes share a common \texttt{\hl{TICKET\_CONTRACT.md}} that codifies status-transition IDs, creation authority by lane, mandatory labels, and~shared schemas.~The~lanes communicate via JSON intermediary files (for example, Lane~2 writes findings to \texttt{\hl{lane\_02\_untracked\_findings.json}}; Lane~3 ingests them and creates Jira tickets). A~design change in any one lane can invalidate assumptions in three to four other~documents.

\textbf{\hl{Representativeness of the Released Synthetic Mini-Specification:}} Reviewers ask how representative the synthetic fixture released in the reproducibility bundle (Appendix~\ref{appendix:c}) is of the real AEGIS surface. We answer with aggregate descriptors of the production specs (Table~\ref{tab:representativeness}), all of which are computed directly from the released anonymized defect catalog or from line- and reference-counts on the production spec files; file-level specification content and Jira surface text are not reproduced in the \hl{Supplementary Materials} %
 because the aggregate descriptors are sufficient for the representativeness comparison, and because the underlying files include third-party data covered by ongoing contractual constraints. The~fixture preserves the property that the audit method targets---cross-document reference density---at the same order of magnitude (real AEGIS 12.77 references per 100 lines; fixture 10.20) while simplifying the scale by roughly thirty-fold (8 files vs.\ 4; median 926 lines per file vs.\ 30). It is a faithful scaled-down reproduction of the cross-document structure that makes cross-document auditing necessary, appropriate for replicating the audit method (as Section~\ref{subsec:cross-llm} does on this exact fixture), but~not a scale model of the production system; absolute defect counts and rounds-to-convergence should not be extrapolated from the fixture to AEGIS. Defect density is intentionally higher in the fixture (5 seeded defects in 114 lines) than in production (per-category rates of 0.58--1.74 per 1000 lines) so that a reviewer can exercise every checklist dimension on a tractable~artifact.

\begin{table}[H]
\caption{\hl{Aggregate} %
 descriptors of the production AEGIS specification surface and the released synthetic mini-specification fixture. AEGIS values computed on the production specs as of 9 June 2026 and on the released 51-row anonymized defect catalog.~Per-category defect rates use the seven-category taxonomy of Section~\ref{subsec:taxonomy}.}
\label{tab:representativeness}

\begin{adjustwidth}{-\extralength}{0cm}

\begin{tabularx}{\fulllength}{lLL}
\toprule
\textbf{Descriptor} & \textbf{Real AEGIS} & \textbf{Synthetic Mini-Spec} \\
\midrule
File count                                & 8 & 4 \\
\midrule
Median lines per file (IQR)               & 926 (696--1098) & 30 (24--32) \\
\midrule
\makecell[l]{Cross-reference density\\(refs/100 lines)}   & 12.77 & 10.20 \\
\midrule
Distinct schema-element count             & 76 & 5 \\
\midrule

\makecell[l]{Per-category defect rate\\(per 1000 lines)}  & version\_drift 1.30; stale\_jira\_ref 1.74; cross\_lane\_schema 0.72; missing\_lane\_coverage 1.01; label\_contract 0.87; semantic\_text 1.16; formula\_timing 0.58 & 5 seeded defects in 114~lines \\
\midrule
Mean files involved per defect            & 1.27 (computed from the released catalog's \texttt{\hl{files\_involved}} column) & 1.40 (computed analogously on the five seeded defects) \\
\midrule
\makecell[l]{Cross-file defect rate\\(\% of defects spanning $>1$ file)}   & 27\% (14 of 51) & 40\% (2 of 5) \\
\midrule
\makecell[l]{Per-file defect distribution\\(count per file, sorted)}  & 12, 12, 11, 9, 9, 5, 4, 3 \linebreak  (sum \mbox{65 file-references} over 51 defects across 8 files) & 2, 1, 1, 1, 0 across 4 files; 2 cross-file defects accounting for the~surplus \\
\bottomrule
\end{tabularx}
\end{adjustwidth}
\noindent{\footnotesize{\hl{Schema-element} %
 count is the distinct backticked field/key identifiers named across all spec files (a broader scope than the four shared JSON contract schemas listed in Table~\ref{tab:metrics}, which count only inter-lane contract objects).}}
\end{table}
\unskip

\begin{table}[H]
\caption{Specification surface metrics for the AEGIS pipeline at the time of the audit. Line counts as reported in the companion paper~\cite{calboreanu2026aegis_companion} (Section~6.1)%
.}
\label{tab:metrics}

\begin{adjustwidth}{-\extralength}{0cm}
\centering 

\begin{tabularx}{\fulllength}{ll}
\toprule
\textbf{Metric} & \textbf{Value} \\
\midrule
Configuration files audited & 8 (seven lane \texttt{\hl{PROMPT.md}} files plus one shared \texttt{\hl{TICKET\_CONTRACT.md}}) \\
Lane PROMPT.md files & 7 (one per lane: Lanes 1--7) \\
Total lane PROMPT lines & 6907 \\
Ticket contract lines & 245 \\
Total specification lines & \textasciitilde7152 \\
Cross-document references & 37 identified (cataloged by grep + manual confirmation across the eight files) \\
Shared data schemas & 4 (\texttt{\hl{fix\_queue}}, \texttt{\hl{untracked\_findings}}, \texttt{\hl{connectivity}}, \texttt{\hl{fallback}}) \\
Shared transition IDs & 4 (To Do, In Progress, On Hold, Done) \\
Ticket-creating lanes & 4 \\
\bottomrule
\end{tabularx}
\end{adjustwidth}
\end{table}

\vspace{-3pt}

AEGIS is also governed by separately documented Context Engineering~\cite{calboreanu2026context}, MANDATE~\cite{calboreanu2026mandate}, LATTICE~\cite{calboreanu2026lattice}, and~TRACE~\cite{calboreanu2026trace} specifications.~These frameworks provide system context, but~the present study evaluates the consistency of the prompt-specification layer rather than validating the frameworks themselves. Specification-surface metrics are summarized in Table~\ref{tab:metrics}.

\subsection{Case Study~Design}

We follow the case-study reporting conventions of Runeson and H{\"o}st~\cite{runeson2009guidelines}.

\textbf{\hl{Case Selection Rationale:}} AEGIS was selected because (a)~it is a production-deployed multi-agent system whose specification surface had grown organically across lane consolidations and architectural changes, (b)~the author had artifact-level access required to run the audit, and~(c)~the system had reached sufficient operational maturity for cross-document audit to surface meaningful~defects.

\textbf{\hl{Unit of Analysis:}} The unit of analysis is the prompt-specification surface (eight files) and the audit process (nine rounds). Convergence is measured at the audit-round level; defect classification is measured at the individual-finding~level.

\textbf{\hl{Propositions:}} RQ1 implies the proposition that \emph{\hl{defect classes exist that single-file review cannot detect by construction}}.~RQ2 implies the proposition that \emph{\hl{iterative auditing exhibits non-monotonic convergence in this case}}.

\textbf{\hl{Chain of Evidence:}} Audit logs (per-round finding lists with file/line citations) are the primary data; the AEGIS specification files are the substrate; the author's classification is the interpretation layer. The~reproducibility bundle (Appendix~\ref{appendix:c}) preserves the audit logs and classification~labels.

\subsection{Audit Protocol (As Used in the Study)}
\label{subsec:audit-protocol}

Each audit round consisted of spawning one or more Claude sub-agents (via the Agent tool) with a checklist covering seven dimensions:

\begin{enumerate}
\item Version consistency within and across files;
\item Cross-lane data contract alignment (field names, schemas);
\item Jira permission boundaries (which lanes may create, comment, and transition);
\item Label conventions per the ticket contract;
\item Lane-count propagation (all references must cite seven lanes);
\item Cadence and scheduler alignment;
\item Internal contradictions between sections of the same document.
\end{enumerate}

Agents read each file top-to-bottom and returned structured findings with exact line numbers. Fixes were applied between rounds using targeted edits, and~verification greps confirmed each fix. Following Weinberg and Freedman~\cite{weinberg1984reviews}, we adopted \emph{\hl{walkthrough}} discipline rather than full Fagan inspection: there was one inspector (the LLM family) rather than the moderator/reader/recorder/inspector role separation that Fagan inspection requires. Section~\ref{subsec:threats} documents the limitations that this~entails.

The audit protocol co-evolved with the fixes. Early rounds used per-file agents; later rounds explicitly loaded producer and consumer specifications together for cross-lane checks.~Section~\ref{subsec:chronology} documents this chronology.~Consequently, the~convergence curve \linebreak  in Section~\ref{sec:results} reflects two interleaved processes---defect remediation and audit refinement---and should be read as describing the joint process rather than as evidence about a fixed audit function. The~final locked protocol used after the study is reproduced in Appendix~\ref{appendix:a}; that locked version is the one that we recommend for~replication.

\subsection{Defect~Coding}

The seven-category taxonomy reported in Section~\ref{subsec:taxonomy} was derived post hoc by a single coder (the author) and is therefore subject to single-coder bias.~This is foregrounded as a central methodological limitation, not a peripheral one. Cemri~et~al.~\cite{cemri2025multiagent_fail} achieved \(\kappa = 0.88\) inter-rater agreement on a comparable taxonomy by using multiple coders; the present study did not. Future replications should run a quick inter-rater check on a defect~subsample.

Category definitions, decision rules for boundary cases, a~worked boundary case, and~one representative example per category are provided in Appendix~\ref{appendix:b}. Categories were treated as mutually exclusive at coding time; any defect that could plausibly fit two categories was assigned to the one matching the \emph{\hl{trigger of detection}}, not the apparent~fix.

\subsection{Convergence Criterion (Operational, Not Mathematical)}

The audit loop terminated when a full-scope audit (all eight files, all seven checklist dimensions) returned zero findings on round 9. Operationally, the~loop resembles audit-to-quiescence: repeat audit-and-fix until a full-scope pass returns no new findings. This is analogous to fixed-point iteration, but the analogy is methodological rather than mathematical: the audit function changes across rounds (Section~\ref{subsec:audit-protocol} caveat), the~LLM auditor is stochastic, and~a single clean pass is not a mathematical fixed point of a stationary~operator.

A stricter protocol would require \emph{\hl{two consecutive clean passes}} to guard against oscillation. We did not adopt that stricter criterion in this study. The~justification for the one-clean-pass rule actually used is operational: the audit was undertaken against a production system on a finite timeline, and~the cost of an additional full-scope round was substantial; we elected to stop at round~9's clean pass and defer the stricter criterion to replication. \emph{\hl{We recommend the two-consecutive-clean-pass rule as the replication criterion}}, and~we specify it formally in Appendix~\ref{appendix:a.3}. This recommendation is foregrounded here rather than buried in the Appendix, because the stopping rule is a known source of premature-termination risk on stochastic LLM auditors~\cite{wang2023self}.

\subsection{Context-Loading~Chronology}
\label{subsec:chronology}

The audit protocol used during the study evolved as documented in Table~\ref{tab:chronology}. The~nine rounds were conducted across approximately five weeks (early April through early May 2026), after~Lane 7 was introduced on 31 March 2026 per the companion paper~\cite{calboreanu2026aegis_companion}. \hl{The~specification surface was held stable across audit rounds, modulo the targeted fixes applied between rounds;} %
 no architectural changes were introduced during the audit~window.

\begin{table}[H]
\caption{Context-loading chronology for the nine audit~rounds.}
\label{tab:chronology}

\begin{adjustwidth}{-\extralength}{0cm}
\centering 

\begin{tabularx}{\fulllength}{cCCc}
\toprule
\textbf{Round} & \textbf{Files Loaded per Call} & \textbf{Cross-Lane Comparison?} & \textbf{Notes} \\
\midrule
1 & 1 & No & Per-file structural~pass \\
2 & 1 & No & Per-file structural~pass \\
3 & 1--2 & Partial & First cross-lane checks for label~conventions \\
4 & 2--3 & Yes & First explicit producer-consumer schema~comparison \\
5 & 2--3 & Yes & Stabilized cross-lane comparison~strategy \\
6 & 8 & Yes & First full-scope~pass \\
7 & 8 & Yes & Full-scope, including changelog and example~data \\
8 & 8 & Yes & Full-scope; surfaced \texttt{\hl{priority\_score}} vs. \texttt{\hl{fix\_priority}} mismatch \\
9 & 8 & Yes & Clean pass; no~findings \\
\bottomrule
\end{tabularx}
\end{adjustwidth}
\end{table}

\vspace{-14pt}

\section{Results}
\label{sec:results}
\unskip

\subsection{Convergence~Data}

\begin{figure}[H]
\includegraphics[width=0.8\textwidth]{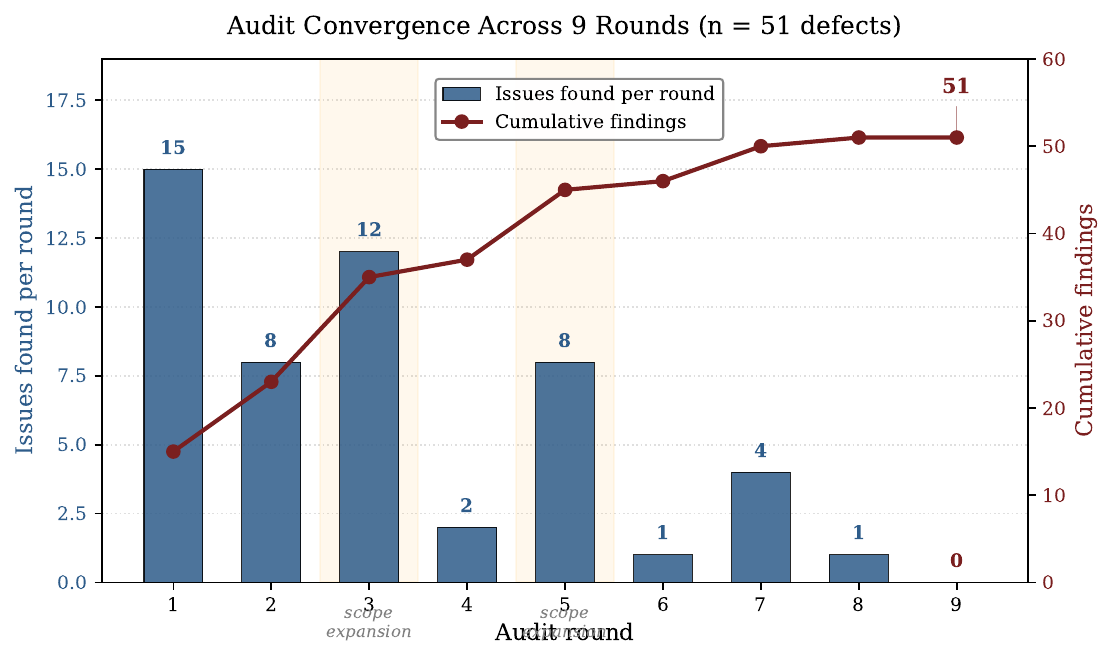}
\caption{\hl{Audit} %
 \hl{convergence} %
 across nine rounds.~Bars show per-round defect counts (left axis); the line shows cumulative totals (right axis). Shaded regions in rounds 3 and 5 indicate observed scope~expansion.}
\label{fig:convergence}
\end{figure}

Table~\ref{tab:convergence} presents the per-round convergence data. The~audit surfaced 51 total prompt-specification defects, distinct from the 51 STRIDE-categorized adversarial code findings reported in~\cite{calboreanu2026aegis_companion} (\hl{Table~4} %
 of the companion). The per-round counts were 15, 8, 12, 2, 8, 1, 4, 1, and~0. Figure~\ref{fig:convergence} shows the convergence curve with per-round discovery rates and cumulative~totals.

\begin{table}[H]
\caption{Audit convergence across nine rounds. Per-round shares are reported with Wilson 95\% confidence intervals computed on the share of the total $n = 51$ defects surfaced in each round. CIs are reported as a descriptive uncertainty band on each round's contribution to the total, not as inference about an underlying defect-generating distribution; round-to-round counts are not independent draws (see Section~\ref{subsec:observations}).}
\label{tab:convergence}
\small
\begin{adjustwidth}{-\extralength}{0cm}
\centering 

\begin{tabularx}{\fulllength}{CCcccc}
\toprule
\textbf{Round} & \textbf{Issues Found} & \textbf{Share} & \textbf{\makecell[c]{95\% CI\\(Wilson)}  } & \textbf{Primary Category} & \textbf{Cumulative} \\
\midrule
1 & 15 & 0.294 & [0.187, 0.430] & Structural: stale versions, removed features, missing lanes & 15 \\
2 & 8  & 0.157 & [0.082, 0.280] & Operational: fallback references, footer versions, lock commands & 23 \\
3 & 12 & 0.235 & [0.140, 0.368] & Semantic: misleading env-var docs, Jira templates, cadence references & 35 \\
4 & 2  & 0.039 & [0.011, 0.132] & Schema: timing assumptions, \texttt{\hl{fix\_queue}} field alignment & 37 \\
5 & 8  & 0.157 & [0.082, 0.280] & Contract: \texttt{\hl{fallback\_enabled}} flag, labels, status contract coverage & 45 \\
6 & 1  & 0.020 & [0.003, 0.103] & Changelog: stale fallback queue reference in version history & 46 \\
7 & 4  & 0.078 & [0.031, 0.185] & Cross-lane: formula mismatch, planned scripts, transition scope & 50 \\
8 & 1  & 0.020 & [0.003, 0.103] & Data contract: \texttt{\hl{priority\_score}} vs. \texttt{\hl{fix\_priority}} field-name break & 51 \\
9 & 0  & 0.000 & [0.000, 0.070] & Clean pass: zero issues across all eight files & 51 \\
\bottomrule
\end{tabularx}
\end{adjustwidth}
\end{table}

\vspace{-18pt}

\unskip

\subsection{Defect~Taxonomy}
\label{subsec:taxonomy}

\begin{figure}[H]
\includegraphics[width=0.95\textwidth]{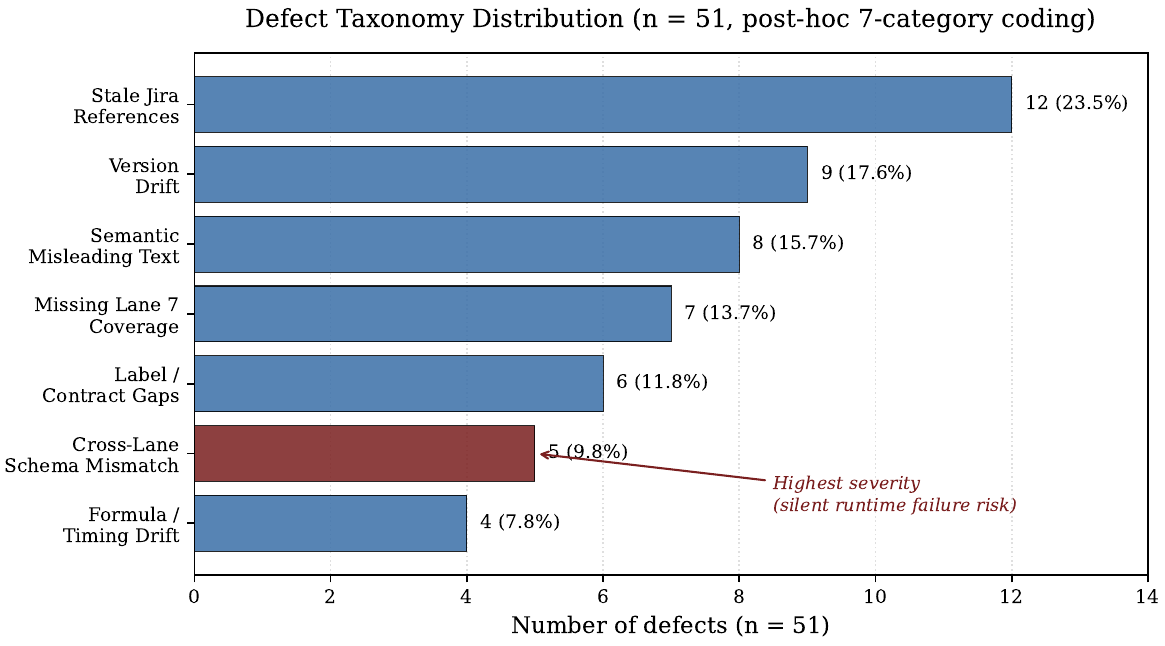}
\caption{Defect taxonomy distribution (n = 51). Cross-lane schema mismatches (highlighted) are the highest-severity category by author~coding.}
\label{fig:taxonomy}
\end{figure}

Table~\ref{tab:taxonomy} presents the post hoc seven-category taxonomy; Figure~\ref{fig:taxonomy} shows the same distribution as a horizontal bar chart with the highest-severity category highlighted. Category definitions, coding rules, and~examples are given in Appendix~\ref{appendix:b}. Stale Jira references (23.5\%) were the most prevalent category, reflecting ripple effects of an earlier architectural change that removed Jira ticket creation from Lane~2. Cross-lane schema mismatches (9.8\%), though~fewer in count, were author-coded as high-severity because such mismatches can cause silent runtime failures (Section~\ref{subsec:observations} and Appendix~\ref{appendix:a}).

\begin{table}[H]
\caption{Defect classification (n = 51). Percentages sum to 99.9 due to~rounding.}
\label{tab:taxonomy}
\begin{tabularx}{\textwidth}{cCC}
\toprule
\textbf{Defect Type} & \textbf{\emph{n}} & \textbf{\%} \\
\midrule
Version drift & 9 & 17.6 \\
Stale Jira references & 12 & 23.5 \\
Cross-lane schema mismatch & 5 & 9.8 \\
Missing Lane 7 coverage (recent extension) & 7 & 13.7 \\
Label/contract gaps & 6 & 11.8 \\
Semantic misleading text & 8 & 15.7 \\
Formula/timing drift & 4 & 7.8 \\
\bottomrule
\end{tabularx}
\end{table}
\unskip

\textbf{\hl{Severity Distribution}} (author-coded per Appendix~\ref{appendix:a.4} rubric): High severity, 5 (9.8\,\%); medium, 32 (62.7\,\%); low, 14 (27.5\,\%). The~high-severity category and the cross-lane schema mismatch category are coextensive in this study: every cross-lane schema mismatch was coded as high-severity per Appendix~\ref{appendix:a.4}'s silent-runtime-failure criterion, and~no other defect class reached the high-severity~threshold.

\textbf{\hl{Objective Severity Proxies:}} The author-coded high/medium/low rubric is supplemented with a lightweight objective metric per category, as reported in Table~\ref{tab:severity-proxies}. These are descriptive surface measurements drawn directly from the released defect catalog and the synthetic mini-spec, not replacements for the rubric. Their purpose is to give a reviewer a category-specific quantity that does not depend on author judgment; the inter-rater reliability check in Section~\ref{subsec:irr} reports $\kappa = 0.46$ for severity, which is independent evidence that severity grading is the most judgment-dependent axis of the taxonomy and worth supplementing in this~way.

\begin{table}[H]
\caption{Objective severity proxies per taxonomy category.~Each proxy is a surface-measurable count that captures one dimension of the defect's blast radius, supplementing (not replacing) the author-coded H/M/L~rubric.}
\label{tab:severity-proxies}

\begin{adjustwidth}{-\extralength}{0cm}

\begin{tabularx}{\fulllength}{lX}
\toprule
\textbf{Category} & \textbf{Proxy} \\
\midrule
version\_drift             & Number of distinct file pairs in which version strings~disagree \\
stale\_jira\_ref           & Number of stale field-name or behavior occurrences~referenced \\
cross\_lane\_schema        & Number of producer--consumer file pairs whose schema fields~disagree \\
missing\_lane\_coverage    & Number of cross-document references that omit one or more~lanes \\
label\_contract            & Number of label-string variants used in place of the contract-canonical~form \\
semantic\_text             & Number of within-file body/changelog pairs whose claims~contradict \\
formula\_timing            & Magnitude of the timing or formula deviation, scaled to the documented~value \\
\bottomrule
\end{tabularx}
\end{adjustwidth}
\end{table}

\vspace{-3pt}

\textbf{\hl{Defect Density:}} Fifty-one defects across 7152 specification lines yielded approximately \mbox{7.1 defects} per thousand lines of specification. This provides an order-of-magnitude reference point, but~direct comparison to code-inspection rates from the Fagan-tradition empirical literature~\cite{fagan1976design,boehm2001defect} is not warranted without comparable inspection-effort normalization: the artifact class (natural-language prompt specifications versus source code), the~inspector (LLM versus human), the~defect definition, and~the inspection depth all~differ.

\subsection{Key~Observations}
\label{subsec:observations}

\textbf{\hl{Observed Non-Monotonic Convergence:}} Rounds 3 and 5 surfaced more issues than round 2. We interpret this as the audit scope expanding rather than a regression: structural fixes in earlier rounds revealed previously masked inconsistencies. Auditor-variance vs. mechanism interpretation is discussed in Section~\ref{subsec:statistical}.

\textbf{\hl{Two Intertwined Signals---Defect-Removal Convergence and Audit-Protocol Maturation:}} The per-round series in Figure~\ref{fig:convergence} blends two processes that the case-study design cannot separate. The~first is \emph{\hl{defect-removal convergence}}: as fixes accumulate between rounds, fewer defects remain to be found, which is the convergence behavior that the protocol is designed to drive. The~second is \emph{\hl{audit-protocol maturation}}: as documented in the context-loading chronology (Table~\ref{tab:chronology}), the~inspector's effective coverage expanded from single-file scope in rounds 1--2, to~partial cross-file scope in rounds 3--5, to~full-scope cross-document loading from round~6 onward. The~bumps at rounds 3 and 5 are therefore consistent with two non-exclusive explanations: structural fixes from earlier rounds unmasking previously hidden inconsistencies (the cascading-edits interpretation of Section~\ref{subsec:cascading}), and~the audit prompt itself becoming capable of finding defect classes that it could not previously see. We interpret the non-monotonic pattern as the joint signature of both processes rather than as an intrinsic property of any one audit method, and~the formal trend statistic reported below is conditioned on this confound. The~locked Appendix~\ref{appendix:a} protocol---with the full seven-dimension checklist and explicit full-scope context loading from round~1---is the version recommended for replicators precisely so that future runs see only the first signal, not the~second.

\textbf{\hl{Cross-Lane Defects Are the Hardest to Catch with Single-File Review:}} The most operationally consequential defect (round~8) was a pre-remediation field-name mismatch: Lane~3 emitted \texttt{\hl{priority\_score}} while Lane~4 was specified to consume \texttt{\hl{fix\_priority}}. This could plausibly have caused a silent runtime failure, such as empty sort results, with~no error signal. The~defect was remediated before the companion paper's final architecture review, and~both lanes now use \texttt{\hl{priority\_score}} (companion paper~\cite{calboreanu2026aegis_companion}, \hl{Section~3.2}%
). By~construction, single-file review cannot detect this class of defect. A~full-scope single-pass review \emph{\hl{might}} detect it; we did not run such a~control.

\textbf{\hl{Specification Aging Outlasts Structural Fixes:}} After all structural issues were resolved, rounds 6 through 8 continued to surface semantic inconsistencies. These were not detected by the grep-based verification used in this study. Parnas~\cite{parnas1994software} identified this phenomenon as software aging; the prompt-specification analog observed here is \emph{\hl{specification aging}}.

\textbf{\hl{Trend Statistic for the Convergence Series.}} A nonparametric Mann--Kendall trend test on the per-round count vector $(15, 8, 12, 2, 8, 1, 4, 1, 0)$ yields $S = -26$, $Z = -2.64$ (with continuity correction; tie-corrected variance $\mathrm{Var}(S) = 90$ from two tied pairs), and~a two-tailed $p = 0.008$, with~Kendall's $\tau = -0.72$. The~series shows a statistically significant decreasing trend by this test. We report this descriptively and treat the $p$-value as approximate for two reasons: rounds are not independent draws from a fixed audit function (the protocol matured across rounds as documented above, and~removed defects do not regenerate), and~the population of detectable defects is bounded by the specification's finite content. The~trend statistic is consistent with the cascading-edits and audit-protocol-maturation interpretation, but it~does not by itself rule out alternatives; see Section~\ref{subsec:statistical}.

\subsection{Defect Category~Progression}

\begin{figure}[H]
\includegraphics[width=0.95\textwidth]{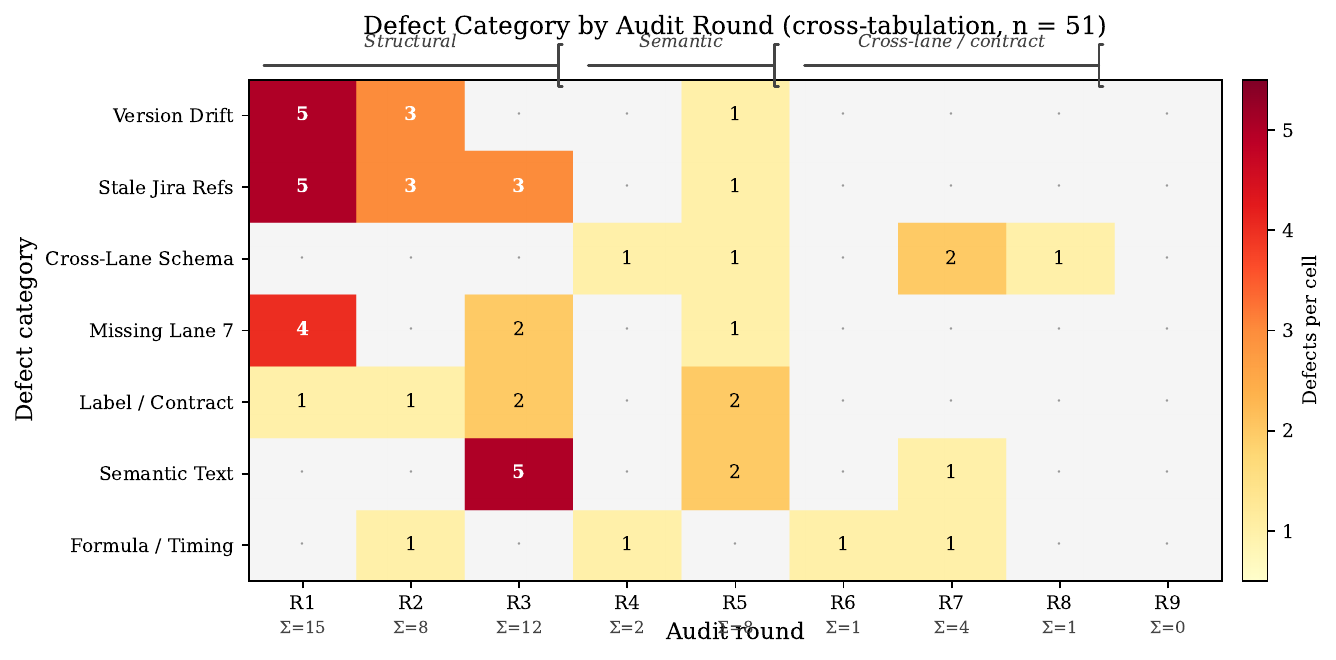}
\caption{\hl{Defect} %
 category by audit round (cross-tabulation, \emph{n} = 51). Phase brackets above the heatmap mark three observed phases of the convergence path: a \emph{\hl{structural}} phase (rounds 1--2) dominated by version drift and stale Jira references, a \emph{\hl{semantic}} phase (round 3) when misleading-text defects peak, and a \emph{\hl{cross-lane/contract}} phase (rounds 4--8) when multi-file context loading first becomes standard and the cross-lane schema and label-contract defects surface. These bracket labels are descriptive of the observed pattern in this case study, not pre-registered phase~definitions.}
\label{fig:progression}
\end{figure}

Figure~\ref{fig:progression} cross-tabulates defect categories against audit rounds; Table~\ref{tab:crosstab} provides full numeric detail.~Three patterns are visible:~First, version drift and stale Jira references concentrate in rounds 1--3 (19 of 21 combined, or~90.5\%), consistent with structural issues surfacing earliest. Second, semantic misleading text peaks in round 3 (5 of 8 total). Third, cross-lane schema mismatches appear exclusively in rounds 4--8 (all five occurrences), consistent with the observation that this class requires multi-file visibility, which (Table~\ref{tab:chronology}) only became standard from round~4.

\begin{table}[H]
\caption{Defect cross-tabulation: category by~round.}
\label{tab:crosstab}
\begin{tabularx}{\textwidth}{Cccccccccccc}
\toprule
\textbf{Defect Type} & \textbf{R1} & \textbf{R2} & \textbf{R3} & \textbf{R4} & \textbf{R5} & \textbf{R6} & \textbf{R7} & \textbf{R8} & \textbf{R9} & \textbf{Total} \\
\midrule
Version drift & 5 & 3 & 0 & 0 & 1 & 0 & 0 & 0 & 0 & 9 \\
Stale Jira refs & 5 & 3 & 3 & 0 & 1 & 0 & 0 & 0 & 0 & 12 \\
Cross-lane schema & 0 & 0 & 0 & 1 & 1 & 0 & 2 & 1 & 0 & 5 \\
Missing Lane 7 & 4 & 0 & 2 & 0 & 1 & 0 & 0 & 0 & 0 & 7 \\
Label/contract & 1 & 1 & 2 & 0 & 2 & 0 & 0 & 0 & 0 & 6 \\
Semantic text & 0 & 0 & 5 & 0 & 2 & 0 & 1 & 0 & 0 & 8 \\
Formula/timing & 0 & 1 & 0 & 1 & 0 & 1 & 1 & 0 & 0 & 4 \\
\midrule
\textbf{Column total} & \textbf{\hl{15} %
} & \textbf{8} & \textbf{12} & \textbf{2} & \textbf{8} & \textbf{1} & \textbf{4} & \textbf{1} & \textbf{0} & \textbf{51} \\
\bottomrule
\end{tabularx}
\end{table}
\unskip

\unskip

\subsection{Cross-LLM Partial Replication on the Synthetic~Mini-Specification}
\label{subsec:cross-llm}

To address the same-LLM-family concern directly---and on a fixture that every reader can re-run---we applied the locked seven-dimension audit protocol (Appendix~\ref{appendix:a}) to the released synthetic mini-specification (Appendix~\ref{appendix:c}): a four-file, $\sim$114-line synthetic intake-triage pipeline (named ATLAS in the released fixture) with five seeded defects spanning five taxonomy categories and known ground truth. Each of four frontier vendors (Anthropic, OpenAI, Google, xAI) received an identical, blinded copy of the mini-spec---with all ground-truth annotations and seeded-defect category names removed---together with the locked checklist and severity rubric, and they~returned a structured finding list under a fixed JSON schema. The~gold-label key was withheld from every model and used only for scoring. Each vendor was run three times to characterize within-model variance, yielding 12 audit traces in total.~Models were called at temperature 0 where supported; models that do not expose a temperature control (the GPT-5.x and Claude-4.x families) were run at their default and the value recorded as such. Exact model identifiers, decoding settings, token counts, and~wall-clock times are recorded in the released run traces (Appendix~\ref{appendix:c}).

\textbf{\hl{Per-Vendor Detection:}} Table~\ref{tab:cross-llm-vendor} reports, per vendor and averaged over three runs, how many of the five seeded defects were detected, the~spurious-finding count, and~agreement with the gold labels on category and severity for detected defects. A~defect is scored as \emph{\hl{detected}} when at least one finding cites the correct file(s) and identifies the underlying issue; following the original study's trigger-of-detection convention, a~correct detection under a different category label still counts as a detection, with~the category disagreement tracked~separately.

\begin{table}[H]
\caption{Per-vendor detection on the synthetic mini-specification (mean of three runs per vendor).}
\label{tab:cross-llm-vendor}
\begin{tabularx}{\textwidth}{cCcccc}
\toprule
\textbf{Vendor} & \textbf{Model} & \textbf{Detected/5} & \textbf{Spurious} & \textbf{\makecell[c]{Cat.\\Agmt}} & \textbf{\makecell[c]{Sev.\\Agmt}} \\
\midrule
OpenAI    & \texttt{\hl{gpt-5.5-2026-04-23}}     & 5.00 & 0.67 & 1.00 & 1.00 \\
Anthropic & \texttt{\hl{claude-opus-4-8}}         & 4.67 & 1.33 & 0.85 & 0.57 \\
Google    & \texttt{\hl{gemini-3.1-pro-preview}}  & 5.00 & 0.00 & 1.00 & 1.00 \\
xAI       & \texttt{\hl{grok-4.3}}                & 4.00 & 0.00 & 1.00 & 1.00 \\
\bottomrule
\end{tabularx}
\end{table}

\vspace{-3pt}

The central result is unambiguous: frontier models that had no access to Claude's authoring process and no ground-truth key independently recovered the seeded defects at high rates. OpenAI's GPT-5.5 and Google's Gemini 3.1 Pro each detected all five seeded defects in every run; the re-baselined Claude detected all five in two of three runs; xAI's Grok detected four of five in every run. Spurious findings were rare (0--1.3 per run on average) and, where they occurred, generally flagged a real but unseeded consistency issue rather than a hallucinated one. This is direct evidence against the shared-blind-spot hypothesis for this fixture: three independent vendors recovered essentially the same defect set that the Claude-family auditor~did.

\textbf{\hl{Per-Defect Detection:}} Table~\ref{tab:cross-llm-defect} gives the per-defect detection rates across the~panel.

\begin{table}[H]
\caption{Per-seeded-defect detection rate (fraction of three runs in which the defect was detected,\linebreak   per vendor).}
\label{tab:cross-llm-defect}
\small
\begin{adjustwidth}{-\extralength}{0cm}
\centering 

\begin{tabularx}{\fulllength}{ccCCCCCCC}
\toprule
\textbf{ID} & \textbf{Category} & \textbf{Sev.} & \textbf{Scope} & \textbf{OpenAI} & \textbf{Claude} & \textbf{Gemini} & \textbf{Grok} & \textbf{Overall} \\
\midrule
\hl{S1} %
 & version\_drift     & low    & single & 1.00 & 1.00 & 1.00 & 1.00 & 1.00 \\
S2 & stale\_jira\_ref   & medium & single & 1.00 & 1.00 & 1.00 & 0.00 & 0.75 \\
S3 & cross\_lane\_schema & high   & cross  & 1.00 & 0.67 & 1.00 & 1.00 & 0.92 \\
S4 & label\_contract    & medium & cross  & 1.00 & 1.00 & 1.00 & 1.00 & 1.00 \\
S5 & semantic\_text     & medium & single & 1.00 & 1.00 & 1.00 & 1.00 & 1.00 \\
\bottomrule
\end{tabularx}
\end{adjustwidth}
\end{table}

\vspace{-3pt}

The two defects that any model missed were the two that this manuscript's argument predicts are hardest. The~cross-lane, highest-severity schema break (S3) was caught by three of four vendors in every run; only Claude missed it, and~only in one of three runs---the run-to-run variance that the three-run protocol (Appendix~\ref{appendix:a}) exists to absorb. The~stale-Jira reference (S2) was caught by three vendors every time but missed by Grok in all three runs---a clean, reproducible, vendor-specific blind spot. The~two blind spots are vendor-specific and non-overlapping, which is itself the case for a multi-vendor panel: no single model is uniformly best, but~the union of the panel detects every defect. A pilot of the Gemini leg on the smaller \texttt{\hl{gemini-2.5-flash}} model missed S3 in all runs, so detection of the cross-lane defect also scales with model capability---a useful bound for replicators choosing models.

\textbf{\hl{Cross-Vendor Agreement:}} Table~\ref{tab:cross-llm-agreement} reports pairwise agreement between vendors on which seeded defects were \emph{\hl{confirmed}} detected (in $\geq$2 of 3 runs): the Jaccard overlap of confirmed-detected sets, and~Cohen's $\kappa$ on the category labels that the two vendors assigned to the defects they both~detected.

\begin{table}[H]
\caption{Pairwise cross-vendor agreement on confirmed~detections.}
\label{tab:cross-llm-agreement}
\small
\begin{tabularx}{\textwidth}{CCCc}
\toprule
\textbf{Vendor A} & \textbf{Vendor B} & \textbf{Jaccard} & \textbf{Cohen's $\kappa$ (Categories)} \\
\midrule
OpenAI & Claude  & 1.00 & 0.76 \\
OpenAI & Gemini  & 1.00 & 1.00 \\
OpenAI & Grok    & 0.80 & 1.00 \\
Claude & Gemini  & 1.00 & 0.76 \\
Claude & Grok    & 0.80 & 0.69 \\
Gemini & Grok    & 0.80 & 1.00 \\
\bottomrule
\end{tabularx}
\end{table}

\vspace{-3pt}

OpenAI, Claude, and~Gemini confirm-detected the identical set of all five defects (Jaccard $= 1.00$ among them); Grok's 0.80 Jaccard against the other three reflects exactly one missing defect (S2), not category disagreement. On~the defects that the vendors co-detected, category agreement is high to perfect ($\kappa = 0.69\text{--}1.00$), with~OpenAI/Gemini/Grok agreeing perfectly pairwise; the only $\kappa$ below 0.76 involves Claude's one divergent category label on the sort-order defect (S5). In~short, the~panel disagrees on at most one borderline category label and one vendor's single coverage gap---not on what the defects~are.

\begin{figure}[H]

\includegraphics[width=0.92\textwidth]{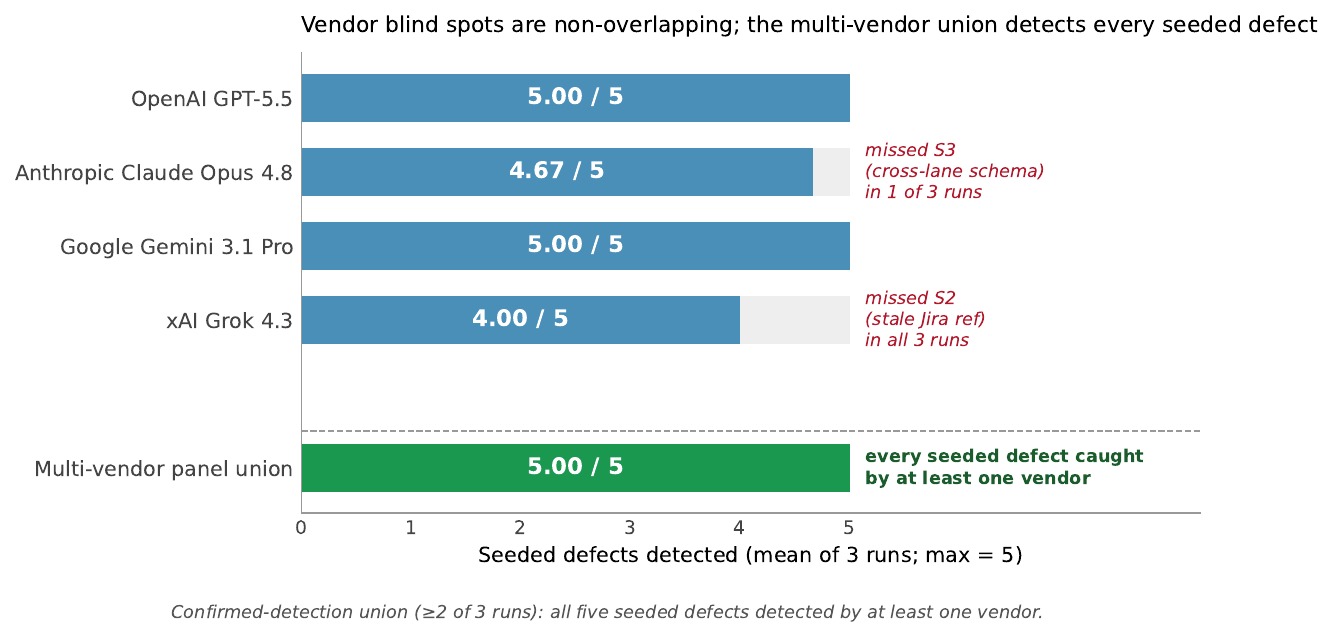}
\caption{Per-vendor detection of the five seeded defects on the synthetic mini-spec (mean of three runs each), with~the multi-vendor panel union shown separately at the bottom. The~two coverage gaps that exist are vendor-specific and non-overlapping (Grok missed the stale-Jira reference S2 in all three runs; Claude missed the cross-lane schema break S3 in one of three runs); the panel union therefore detects every seeded defect. This is direct evidence against the shared-LLM-family-blind-spot hypothesis on this fixture: if the concern were operative, the~non-Claude vendors should miss the same defects that Claude misses, which they do~not.}
\label{fig:cross-llm-detection}
\end{figure}

\textbf{\hl{Scope:}} This is a partial replication, bounded in three specific ways.~First, it is not a multi-\emph{\hl{system}} replication: running the protocol on additional independent specification surfaces remains future work, and~the cross-vendor agreement reported here does not bound the same-LLM-family risk on the production AEGIS specs, only on the released fixture. Second, it is not a single-pass full-scope control comparison; the panel does not compare iterative auditing against a one-pass alternative, only against itself across vendors.~Third, it is not a human-only or hybrid human--LLM comparison; no human reviewer audited the mini-spec. What the panel does provide---and the specific scope under which it should be cited---is direct empirical evidence that non-Claude frontier models recover the same seeded defects under the locked protocol on a public fixture, which is the specific evidence that the shared-blind-spot concern calls for. Figure~\ref{fig:cross-llm-detection} visualizes this as a 5 $\times$ 4 detection matrix:~18 of 20 cells at 1.00, and~the two coverage gaps that exist are vendor-specific and non-overlapping. All twelve run traces, the~gold-label key (released after the v15 review cycle to preserve the held-out replication value), and~the scoring scripts can be found in the \hl{Supplementary~Materials}%
.

\vspace{-2pt}

\subsection{Inter-Rater Reliability: Partial~Replication}
\label{subsec:irr}

The seven-category taxonomy and severity grading reported above were developed and applied by a single coder (the author). Following the inter-rater reliability conventions of empirical software engineering~\cite{cemri2025multiagent_fail}---noting that Cemri~et~al.~\cite{cemri2025multiagent_fail}%
's reported $\kappa = 0.88$ benchmark is on a different 14-category multi-agent-failure taxonomy with multiple human coders, so it serves as a reference point for the methodology rather than a direct comparison---we report a partial replication on a stratified subsample of the released 51-defect catalog (Appendix~\ref{appendix:c}).

\textbf{Design:} We drew a stratified random subsample of 18 of the 51 cataloged defects (reproducible; fixed random seed 42), constrained to include at least one defect from each of the seven taxonomy categories. A~second coder then re-coded each defect's category and severity from the locked taxonomy rules and severity rubric alone, blind to the original author labels (the second coder received only the redacted defect description and the files involved). We report Cohen's $\kappa$ between the original author labels and the second-coder labels, on~category and on severity, each with a nonparametric bootstrap 95\% confidence interval (5000 resamples of the 18-defect subsample), along with~raw percent~agreement.

\textbf{Second Coder:} For the second coder, we used the same four-vendor LLM panel as in Section~\ref{subsec:cross-llm}; each model independently coded the subsample from descriptions alone, and~the modal label across the panel served as the second coder. We frame this honestly as multi-LLM coder agreement on a held-out subsample---evidence that the taxonomy categories are model-distinguishable from descriptions alone---and not as a substitute for human inter-rater reliability. A~human second-coder replication is the natural next step and is supported by the same instrument; the pipeline supports re-running the analysis with a human coder's CSV in place of the LLM-panel modal~labels.

\textbf{Results:} Table~\ref{tab:irr} reports the agreement~statistics.

\begin{table}[H]
\caption{Inter-rater reliability on the 18-defect stratified subsample. Second coder: four-LLM panel modal consensus. The~four panel coder identifiers (each model called once per defect to produce a label; the modal label across the four served as the second coder) are OpenAI \texttt{gpt-5.5-2026-04-23}, Anthropic \texttt{claude-opus-4-8}, Google \texttt{gemini-2.5-flash}, and xAI \texttt{grok-4.3}.~The~Gemini coder ran at flash tier for this from-description categorization task because category-from-description is less capability-sensitive than the Section~\ref{subsec:cross-llm} detection audit, which used \texttt{gemini-3.1-pro-preview}; all four model identifiers are recorded per coding decision in the released panel-coding detail (\texttt{irr\_panel\_detail.json}).}
\label{tab:irr}
\begin{tabularx}{\textwidth}{CccC}
\toprule
\textbf{Metric} & \textbf{Value} & \textbf{95\% CI} & \textbf{Landis--Koch Band} \\
\midrule
Cohen's $\kappa$---category   & 0.801 & [0.564, 1.000] & almost~perfect \\
Cohen's $\kappa$---severity   & 0.458 & [0.008, 0.811] & moderate \\
\% agreement---category       & 0.833 &---& raw~agreement \\
\% agreement---severity       & 0.722 & --- & raw~agreement \\
\bottomrule
\end{tabularx}
\end{table}

\vspace{-3pt}

Category agreement is substantial-to-almost-perfect ($\kappa = 0.80$), indicating that the seven categories are applied consistently from descriptions alone across four independent vendors.~Severity agreement is moderate ($\kappa = 0.46$), but~the 95\% confidence interval [0.008, 0.811] spans from near-chance to almost-perfect: the lower bound approaches the ``slight'' Landis--Koch band, so the moderate point estimate should be read as the central tendency of a wide interval rather than as a precise reliability claim.~This is consistent with severity grading being the most judgment-dependent dimension of the taxonomy, and it~is the empirical motivation for the objective severity proxies in Table~\ref{tab:severity-proxies}. The~severity $\kappa$ is unweighted (nominal); a linear-weighted $\kappa$ that credits adjacent-level near-misses (e.g., medium $\leftrightarrow$ high) would be~higher.

\textbf{Disagreement Audit:} The category disagreements are interpretable rather than arbitrary, and~they cluster on one taxonomy boundary. All three on this subsample re-coded the author's label as \texttt{semantic\_text}: two \texttt{formula\_timing} defects (a hard-stop budget quoted inconsistently in an operator-facing example, three of four coders; a changelog entry referencing a removed file, two of four) and one \texttt{stale\_jira\_ref} defect (a relocated ``fallback queue'' diagnostic, on~which all four coders agreed). Each turns on a trigger-of-detection judgment---which checklist dimension surfaced the defect---rather than a category being indefensible, and~the recurring drift into \texttt{semantic\_text} (internal contradiction) is a concrete, actionable signal for sharpening the \texttt{formula\_timing} and \texttt{stale\_jira\_ref} definitions against~it.

\textbf{Scope:} This is a partial replication on a subsample, with~an LLM-panel second coder; it bounds, rather than eliminates, the~single-coder threat. The~taxonomy claim is refined \hl{from ``post hoc, single-coded'' to ``post hoc, single-coded with $\kappa$-validated partial replication.''} %
 The subsample, both coders' labels, the~$\kappa$ computation, and~the bootstrap CIs are included in the Supplementary Materials for independent~recomputation.

\section{Discussion}
\label{sec:discussion}
\unskip

\subsection{Three Distinct Notions of~Review}

``Single-pass review'' is ambiguous. We use three distinct terms:

\begin{itemize}
\item \textbf{Single-File Review:} The auditor reads one specification file at a time, without explicit cross-file comparison.~By~construction, single-file review cannot detect cross-file schema mismatches.
\item \textbf{Initial Full-Scope Pass:}~The auditor reads all files in one context window, once, with~cross-file comparison enabled. We did not run this as a control.
\item \textbf{Iterative Full-Scope Auditing:} Repeated full-scope passes with audit-and-fix between rounds, until~quiescence. This is the protocol distilled from the case study.
\end{itemize}

The defensible empirical claim of this paper is that, in~this system, single-file review missed cross-document defect classes by construction, and~one-and-done audit (any single-pass design) left later-discovered defects unresolved under the case-study protocol. We do not claim that no full-scope single-pass design could in principle have surfaced~them.

\subsection{Cascading~Edits}
\label{subsec:cascading}

Fixing issue A in round N can create a new inconsistency B that was masked by A. For~example, removing \texttt{fallback\_enabled: true} from Lane~2's config (round 5) made the changelog entry referencing \texttt{jira\_fallback\_queue.md} newly incorrect (caught in round 6). This is analogous to regression introduction in traditional code~review.

\subsection{Cross-Document~Visibility}

No single-file audit can detect that Lane~4 expects \texttt{fix\_priority} while Lane~3 emits \texttt{priority\_score}. This requires reading both files in the same pass and comparing schemas field-by-field.~The~audit protocol itself was iteratively refined to support this, as~documented in Table~\ref{tab:chronology}; convergence in this case study therefore depends on both fixing defects and improving the audit~prompt.

\subsection{Implications for Prompt Engineering Practice (Case-Bounded)}

The following are observations from this single case study under the locked protocol, not generalizations to other systems, scales, or~LLM families.~Each item is framed as \emph{what we observed in this system}, not as a normative prescription.~Replication across diverse architectures with dissimilar models is required before any of the points below can be treated as practice~recommendations.

\begin{itemize}
\item \textbf{Treat Prompt Specifications as Code, in~this System:} The behavior that we observed---prompt files carrying data contracts and integration logic, and~breaking them silently when a single document was edited in isolation---is consistent with the promptware position~\cite{chen2025promptware}.~Whether this generalizes to LLM-orchestrated pipelines of different scales or architectures is an open question.
\item \textbf{Iterative Auditing Budget at this Scale and Protocol:} In this case, the locked protocol required nine rounds on a 7152-line, eight-document surface to reach a clean pass; the cross-document defects in rounds 7--8 would have been missed by earlier rounds. The~metric that we used was a full-scope clean pass, not a fixed round count. Rounds-to-convergence at other scales is unknown.
\item \textbf{Cross-Document Audits Caught Contract Defects Single-File Review Could Not, in~this Study:} By construction, single-file review cannot detect cross-lane data-contract issues; in this study every cross-lane defect surfaced only after multi-file context loading was enabled. A~full-scope single-pass review might or might not detect them; we did not run such a control.
\item \textbf{Verification Greps Were Necessary but Not Sufficient, in~this Case:}~Automated string matching caught structural issues but missed semantic drift on this system's specification surface.
\end{itemize}

The cross-LLM partial replication in Section~\ref{subsec:cross-llm} addresses the case-bounded character of these claims from one direction---different vendors recover the same seeded defects under the locked protocol on a public fixture---but does not address multi-\emph{system} generalization, which remains an open topic for future work. We also note explicitly here what the cross-LLM panel did \emph{not} provide: a single-pass full-scope control comparison and a human-only or hybrid human--LLM review.~The~single-pass control would have required running an additional cohort of full-scope audits on the same fixture under a different stopping rule; the human-only and hybrid comparisons would have required coder recruitment outside the revision window. Each remains the natural next study; the cross-LLM panel is the alternative that we could execute on the released public fixture in this revision~cycle.

\textbf{Operational Outcomes Are Out-of-Scope and Reported Elsewhere:}~This paper characterizes the audit process and its convergence behavior; it does not measure whether correcting the 51 surfaced defects improved system performance, reliability, or~operational outcomes. Post-fix runtime behavior of the same AEGIS pipeline (STRIDE-categorized adversarial code findings, lock-handling and race-condition fixes, FMEA safety analysis) is reported in the companion preprint~\cite{calboreanu2026aegis_companion}; the 51 specification defects audited here are distinct in source, method, and~content from the 51 code findings reported there.~A~full operational-validation study that ties specific spec-fix events to downstream runtime metrics is the natural follow-on and is left for future~work.

\subsection{Threats to~Validity}
\label{subsec:threats}

We identified four categories of threats following Runeson and H{\"o}st~\cite{runeson2009guidelines} and Wohlin et al.~\cite{wohlin2012experimentation}. The~first two---shared LLM blind spots and single-coder taxonomy---were the most-flagged concerns in peer review and are foregrounded here; both are now partially bounded by the v15 evidence in Sections~\ref{subsec:cross-llm} and~\ref{subsec:irr}, but not eliminated by~it.

\subsubsection{Same-LLM-Family Auditing and Single-Coder~Taxonomy}
\label{subsec:threat-family-coder}

Two related limitations sit at the foundation of this case study and motivated the v15 partial~replications.

\textbf{Same LLM Family Writes and Audits:} The same LLM family (Claude) both authored AEGIS's prompt specifications and ran the audit; the inspector and the author could in principle share blind spots, and~a clean round-9 pass demonstrates only internal consistency under that single family's vantage point, not ground-truth correctness. The~cross-LLM panel in Section~\ref{subsec:cross-llm} bounds this concern empirically on the released fixture: three independent vendors (OpenAI, Google, xAI) recovered the same seeded defects under the locked protocol, and~the panel's blind spots are vendor-specific and non-overlapping, so the multi-vendor union detected every defect. This is partial replication on a public fixture, not a multi-vendor audit of the production AEGIS specs themselves; the same-family risk for AEGIS proper is reduced, not~eliminated.

\textbf{Single-Coder Taxonomy:} The seven-category taxonomy was developed and applied post hoc by a single coder (the author). Section~\ref{subsec:irr} reports an inter-rater reliability partial replication on a stratified 18-defect subsample with a four-LLM panel as a modal second coder: Cohen's $\kappa$ on category is 0.80 (almost perfect), while on~severity it is 0.46 (moderate). Category agreement is high, and severity less so---consistent with severity being the more judgment-dependent axis. We frame the LLM-panel coder honestly as multi-LLM coder agreement on a held-out subsample, not as a substitute for human inter-rater reliability; a human second coder remains the strongest replication and is supported by the same~instrument.

The taxonomy claim in this paper is therefore positioned as a post hoc categorization with $\kappa$-validated partial replication, not as a fully validated empirical-SE~taxonomy.

\subsubsection{External Validity and~Representativeness}

This study examines a single system ($n = 1$) without a control group; the seven-category taxonomy is post hoc and may not be exhaustive. The~implication is direct: claims in Section~\ref{sec:discussion} should be read as case-specific. The~representativeness analysis in Section~\ref{sec:methods} compares aggregate descriptors of the production AEGIS surface against the released synthetic mini-spec (Table~\ref{tab:representativeness}); the fixture preserves cross-reference density at the same order of magnitude while simplifying scale by roughly thirty-fold, which licenses replicating the audit \emph{method} but not extrapolating absolute counts. Multi-system replication on independent specification surfaces remains the strongest answer and is left for future~work.

\subsubsection{Construct and Statistical~Validity}
\label{subsec:statistical}

\textbf{Construct:} Convergence to zero findings does not establish correctness, only internal consistency under the locked protocol's coverage. Runtime validation remains a necessary complementary quality gate; we report nothing about runtime behaviour~here.

\textbf{Statistical and Conclusion Validity:} The sample sizes are modest ($n = 9$ rounds, \mbox{$n = 51$ defects}). The~Mann--Kendall trend test reported in Section~\ref{subsec:observations} is significant ($p = 0.008$), but the rounds are not independent draws (audit-scope expansion is the explicit confound, and~removed defects do not regenerate). With~a stochastic LLM auditor and no repeat trials per round, the~bumps at rounds 3 and 5 cannot be statistically distinguished from auditor variance without a control comparison; the cascading-edits interpretation is one explanation among several. Following Wohlin~et~al.~\cite{wohlin2012experimentation}, conclusion validity concerns the treatment--outcome relationship: with one stochastic auditor, a~co-evolving protocol, and~no repeat-trial design, alternative explanations for the observed convergence pattern cannot be ruled out. Future replications should fix the protocol before remediation begins and run repeat-trial passes per~round.

\subsubsection{Audit-Protocol Co-Evolution and Sequence~Effects}

The audit protocol co-evolved with the discovered defects (Sections~\ref{subsec:audit-protocol} and \ref{subsec:observations}). Later prompts may have become tuned to the known defect space, weakening the convergence interpretation. Related, later rounds operated on artifacts already changed by prior fixes, so the order in which fixes were applied may have shaped the observed category progression: a different fix order might have surfaced different categories at different rounds. The~locked Appendix~\ref{appendix:a} protocol was finalized after the study and should be applied as-is in future replications to address both confounds; the recommended two-consecutive-clean-pass stopping rule, promoted from the Appendix into the main text (Section~\ref{sec:methods}, Convergence Criterion), is one such~hardening.

\section{Conclusions}
\label{sec:conclusions}

This paper presents a single-system empirical case study of iterative, agent-driven auditing of prompt specifications in a production seven-lane orchestration pipeline. Across nine rounds, 51 prompt-specification defects were surfaced and remediated. The~audit terminated when round~9 returned zero findings under the case-study protocol's one-clean-pass stopping rule; we recommend a stricter two-consecutive-clean-pass criterion for~replication.

We make four case-bounded claims: First, in~this case study, single-file review missed cross-document defect classes by construction. Second, the~observed convergence path was non-monotonic, consistent with cascading edits and audit-scope expansion; a Mann--Kendall trend test on the per-round counts was significant ($p = 0.008$, Kendall's $\tau = -0.72$) but conditioned on the round-non-independence confound, so we report the test descriptively. Third, the~audit protocol itself co-evolved with the fixes; the locked, recommended version is reproduced in Appendix~\ref{appendix:a}. Fourth, two partial replications on a released public synthetic mini-specification bound the most-flagged threats to validity: a four-vendor cross-LLM panel (12 traces; OpenAI, Anthropic, Google, and~xAI) detected the seeded defects independently of Claude, with~the multi-vendor union recovering all five; and an inter-rater reliability check on a stratified 18-defect subsample of the catalog returned Cohen's $\kappa = 0.80$ on category (almost perfect) and $\kappa = 0.46$ on severity (moderate).

Iterative full-scope auditing is a candidate practice for LLM-orchestrated systems where prompt specifications serve as the behavioral contract between autonomous agents. The~v15 partial replications bound the same-LLM-family and single-coder threats on a public fixture but do not address multi-\emph{system} replication; the latter remains required, alongside a single-pass full-scope control and a human second coder, to~validate~generalizability.

\vspace{+6pt}

\supplementary{{The following supporting information can be downloaded at \linksupplementary{s1}: the reproducibility bundle described in the Data Availability Statement, comprising per-round defect counts (\texttt{round\_counts.csv}), the anonymized 51-row defect catalog (\texttt{defect\_catalog.csv}), the locked audit checklist, prompt template, and severity rubric, the context-loading chronology, the synthetic mini-specification with five seeded defects, the twelve cross-LLM audit traces with detection and agreement CSVs, the inter-rater reliability subsample and Cohen's $\kappa$ computation, the aggregate AEGIS representativeness descriptors, and a \texttt{README.md} with replication instructions. %
}}

\funding{This research received no external~funding.}

\institutionalreview{{\textls[+20]{Not applicable.~This~study did not involve human or animal subjects.}}}

\informedconsent{Not applicable.}

\dataavailability{A reproducibility bundle accompanies this submission as Supplementary Materials containing the following: per-round CSVs; the anonymized 51-row defect catalog with category, severity, files-involved, and~dimension metadata per defect; the locked checklist, prompt template, and~severity rubric; the context-loading chronology; the synthetic mini-specification for protocol validation; the 12 cross-LLM audit traces (Section~\ref{subsec:cross-llm}) with detection/agreement CSVs; the inter-rater reliability subsample, panel detail, and~$\kappa$ computation (Section~\ref{subsec:irr}); aggregate AEGIS representativeness descriptors (Table~\ref{tab:representativeness}); and the multi-vendor evaluation pipeline used to produce the new evidence. The~gold-label key for the mini-specification is held back from public release until after the v15 review cycle to preserve the fixture's held-out replication value; reviewers may request it from the authors. The~full AEGIS specification files themselves and the Jira backlog text are not included in the Supplementary Materials because the aggregate descriptors in Table~\ref{tab:representativeness} are sufficient for the representativeness comparison, and~because the underlying files include third-party data covered by ongoing contractual constraints with customers of The Swift Group. The~bundle is sufficient for full replication of the audit method against any independent specification surface.}

\acknowledgments{During the preparation of this study, the~author used Anthropic's Claude (model versions: Claude Opus and Claude Sonnet, active across March--May 2026) as the inspector-agent in the case-study protocol described in the Materials and Methods section and the audit protocol Appendix.~The~use of these large language models is the substantive empirical subject and methodology of the study, rather than a writing-assistance tool. The~author has reviewed all model-generated outputs, performed all defect coding, constructed the taxonomy, and~takes full responsibility for the content of this publication. No generative AI tool was used for superficial text editing beyond the documented~methodology.}

\conflictsofinterest{The author is employed by The Swift Group, LLC, which holds commercial licensing rights to products built on the AEGIS reference implementation. The~audit method and data presented here are released independently of any commercial product. The~employer had no role in the study design, the~audit protocol design, the~analysis, the~interpretation, or~the decision to~submit.}

\abbreviations{Abbreviations}{The following abbreviations and notations are used in this manuscript:\\

\vspace{-3pt}

\noindent
\begin{tabular}{@{}ll}
AEGIS & Autonomous Engineering Governance and Intelligence System \\
ATLAS & Synthetic four-lane intake-triage pipeline used as the released mini-spec \\
CI & Confidence interval \\
FMEA & Failure Mode and Effects Analysis \\
IQR & Interquartile range \\
IRR & Inter-rater reliability \\
JSON & JavaScript Object Notation \\
$\kappa$ & Cohen's kappa (chance-corrected inter-rater agreement) \\
LLM & Large language model \\
MCP & Model Context Protocol \\
MLT & MANDATE/LATTICE/TRACE governance stack \\
QA & Quality assurance \\
RQ & Research question \\
SE & Software engineering \\
\multirow{2}{*}{STRIDE} & Spoofing, Tampering, Repudiation, Information disclosure, Denial of service, Elevation\\
& of privilege \\
$S$ & Mann--Kendall statistic (sum of signed pairwise comparisons) \\
$\tau$ & Kendall's tau (rank-correlation effect size) \\
$Z$ & Standard normal test statistic (Mann--Kendall) \\
\end{tabular}}

\appendixtitles{yes} 
\appendixstart
\appendix

\section{Locked Audit Checklist, Prompt Template, and Method~Specifics}
\label{appendix:a}
\unskip

\subsection{Locked Checklist (Seven Dimensions)}\label{appendix:a.1}

The full locked checklist is reproduced in the Supplementary File \texttt{audit\_checklist}\linebreak  \texttt{\_locked.md}.~The~seven dimensions are version consistency, cross-lane data contract alignment, Jira permission boundaries, label conventions, lane-count propagation, cadence and scheduler alignment, and internal~contradiction.

\subsection{Locked Prompt~Template}\label{appendix:a.2}

The full locked prompt template is reproduced in the Supplementary File \texttt{audit\_\allowbreak prompt\_\allowbreak template.md}. It specifies the structured-output schema (id, file, line, dimension, category, severity, description, suggested\_fix, uncertainty), the~trigger-of-detection rule for category assignment, and~the severity~rubric.

\subsection{Recommended Convergence~Criterion}\label{appendix:a.3}

Terminate when two consecutive full-scope passes return zero findings. This is stricter than the criterion used in the case study, which required only one clean pass.

\subsection{Severity~Rubric}\label{appendix:a.4}

The full severity rubric is reproduced in the Supplementary File \texttt{severity\_rubric.md}. High severity covers silent runtime failure risk and producer--consumer schema breaks; medium covers documented-behavior errors and operator-misleading text; low covers cosmetic and internal-doc~drift.

\subsection{Method Specifics for LLM-Based~Auditing}\label{appendix:a.5}

For replication, report the following: LLM family (Claude, this study), specific model versions (mix of Claude Opus and Sonnet across nine rounds), decoding parameters (default temperature; no custom sampling), context-window policy (single-file rounds 1--2, expanding to multi-file rounds 3--5, full-scope rounds 6--9), sub-agents per round (1--3), and tooling (Claude Agent tool with file-read access; no cached memories; fresh session per round). A~more rigorous replication would lock the model version and decoding parameters before remediation~begins.

\section{Defect Coding Rules and~Examples}
\label{appendix:b}

\textbf{\hl{Coder:}%
} Single coder (the author).~Inter-rater validation was not performed; this is acknowledged as a central limitation in Section~\ref{subsec:threats}.

\textbf{\hl{Mutual Exclusivity:}} Categories are mutually exclusive. Boundary cases were assigned to the category matching the trigger of detection, not the apparent~fix.

\textbf{Worked Boundary Case:}~A defect surfaced in round 5 read ``Lane 5's diagnostic checklist instructs the operator to confirm that the \texttt{fallback\_enabled} flag is set to \texttt{true} if Lane~2 is in degraded mode, but~Lane~2's config file has \texttt{fallback\_enabled: false} after the prior architectural change.'' This defect could plausibly be coded as either \texttt{semantic\_text} (the diagnostic instruction has become misleading) or \texttt{label\_contract} (the flag-value contract has drifted). The~trigger of detection was a label-contract dimension check (dimension 4), so it was coded \texttt{label\_contract}.

The seven category definitions and one example per category are reproduced in the Supplementary File \texttt{audit\_checklist\_locked.md} and the manuscript's full markdown~source.

\section{Reproducibility~Artifact}
\label{appendix:c}

{\sloppy The reproducibility bundle (Supplementary Materials) contains the following: \texttt{round\_counts.csv}, \texttt{defect\_catalog.csv} (anonymized 51-row catalog with category, severity, files-involved, and~dimension per defect), \texttt{audit\_\allowbreak checklist\_\allowbreak locked.md}, \texttt{audit\_\allowbreak prompt\_\allowbreak template.md},\linebreak   \texttt{severity\_rubric.md}, \texttt{context\_\allowbreak loading\_\allowbreak chronology.csv}, \texttt{synthetic\_mini\_specificati}\linebreak \texttt{ on.md} (a four-file mini-specification with five seeded defects across five taxonomy categories for protocol validation against ground truth), \texttt{aegis\_\allowbreak aggregate\_\allowbreak descriptors.md} (the aggregate descriptors reported in Table~\ref{tab:representativeness}), and~a \texttt{README.md} with replication instructions.~The~full AEGIS specification files and Jira backlog are not included because the aggregate descriptors and the released catalog are sufficient for the representativeness analysis, and because the files contain customer-attached third-party data; the bundle is sufficient to replicate the audit \emph{method} against any independent specification~surface.\par}

\newpage
\begin{adjustwidth}{-\extralength}{0cm}
\reftitle{References}

\PublishersNote{}
\end{adjustwidth}

\end{document}